\documentclass[a4paper]{article}
\usepackage{Odyssey2022}
\usepackage{epsfig,amssymb,amsmath}
\usepackage{multirow}
\usepackage{booktabs}

\ninept

\title{C-P Map: A Novel Evaluation Toolkit for Speaker Verification}

\name{Lantian Li, Di Wang, Wenqiang Du, Dong Wang
\thanks{
This work was supported by the National Natural Science Foundation of China (NSFC) under Grants No.62171250,
and also the Tsinghua-SPD Bank Joint Research Center for Digital Finance Technologies.
Dong Wang is the corresponding author.}
}
\address{Center for Speech and Language Technologies, Tsinghua University \\
{\small \tt \{lilt,wangdi,duwq\}@cslt.org, wangdong99@mails.tsinghua.edu.cn}}
\begin{document}
\maketitle

\begin{abstract}

Evaluation trials are used to probe performance of automatic speaker verification (ASV) systems. In spite of the
clear importance and impact, evaluation trials have not been seriously treated in research and engineering practice.
This paper firstly presents a theoretical analysis on evaluation trials and highlights potential bias
with the most popular cross-pairing approach used in trials design.
To interpret and settle this problem, we define the concept of trial config and C-P map derived from it.
The C-P map measures the performance of an ASV system on various trial configs in a 2-dimensional map.
On the map, each location represents a particular trial config and its corresponding color represents the system performance.
Experiments conducted on representative ASV systems show that the proposed C-P map
offers a powerful evaluation toolkit for ASV performance analysis and comparison.
The source code for C-P map has been release at \emph{https://gitlab.com/csltstu/sunine}.

\end{abstract}

\section{Introduction}
\label{sec:intro}

Automatic speaker verification (ASV) aims to testify the claimed speaker identity of a speech  signal~\cite{campbell1997speaker,reynolds2002overview,kinnunen2010overview,hansen2015speaker,bai2021speaker}.
After decades of research, current ASV systems have achieved fairly good performance, at least in benchmark evaluations.
For example, with a bunch of state-of-the-art architectures/techniques, such as ResNet-based topology~\cite{he2016deep,desplanques2020ecapa,zhou2021resnext},
self-attentive pooling~\cite{zhu2018self}, angular margin loss~\cite{wang2018additive,deng2019arcface} and
score normalization \& calibration~\cite{kenny2010bayesian,matejka2017analysis},
researchers have reported equal error rates (EER) less than 1.0\% on the VoxCeleb dataset, partly due to
VoxSRC~\cite{chung2019voxsrc,nagrani2020voxsrc}, one of the most popular speaker verification challenges.
Similar low EERs were also reported on SITW~\cite{mclaren2016speakers}, another famous benchmark dataset.

In spite of the impressive EER results in benchmark evaluations, the genuine performance of
modern ASV systems is dubious in real deployment applications.
Practitioners often observe substantial performance gap between benchmark test and experience of end users,
which we call \emph{benchmark-deployment gap}.

To interpret and settle the benchmark-deployment gap, numerous studies have been carried out.
Most of the research focuses on the \emph{data theme}, which hypothesizes that the performance gap
is largely attributed to acoustic mismatch.
Based on this belief, various benchmark datasets were elaborately designed to simulate real-life acoustic conditions.
For example, HI-MIA~\cite{qin2020hi} was designed for near-far field mismatch,
NIST SRE~\cite{martin2009nist,sadjadi20172016} involved long-short mismatch and channel mismatch.
Interestingly, even with the acoustic mismatch, researchers found that good performance can still be attained on benchmark tests.
A possible reason is that these elaborately designed benchmarks, although involve some types of acoustic mismatch, yet
 reflect the true complexity of real-life conditions.
Recently, a more challenging dataset CN-Celeb~\cite{fan2020cn,li2022cn} was released, with an intentional design to involve more real-life complexity,
in particular multi-genre and cross-genre phenomena. Experiments on CN-Celeb demonstrated much higher EER results than on other benchmarks.
Solving real-life data complexity remains a hot and challenging topic.

Besides the data complexity, there is another issue that seems equally important but less attended: the bias on evaluation trials.
It is well known that trials are used to probe and measure performance of ASV systems.
Therefore, if the trials are not appropriately designed, the performance of the target system cannot be correctly measured.
Unfortunately, in most existing benchmark tests, the trials are designed simply based on cross-pairing.
This trivial design leads to trials that might be quite different from those encountered in real-life scenarios.
In particular, cross-pairing tends to produce a large proportion of \emph{easy} trials,
leading to over optimistic performance estimation.
We argue that the \emph{trial bias} problem is a primary source for the benchmark-deployment gap.

In this paper, we firstly conduct a theoretical analysis on evaluation trials and highlight the potential trial bias problem.
Our analysis shows that an ASV system may exhibit very different performance when probed by different settings of trials,
and therefore a prudent performance evaluation should involve multiple and diverse trial settings.
We call each setting of trials a \emph{trial config}, and the performance with
various trial configs form a 2-dimensional map, named \emph{Config-Performance (C-P) map}.
We will show that C-P map is a powerful tool for system analysis, tuning and comparison.
For instance, the benchmark-deployment gap can be interpreted as the performance discrepancy between two different trial configs on the C-P map.

The rest of the paper is organized as follows.
Section~\ref{sec:bias} presents a theoretical analysis on evaluation trials and discuss the trial bias problem,
and Section~\ref{sec:map} presents the concept of trial config and C-P map derived from it.
Section~\ref{sec:exp} presents the experimental results with C-P map of several representative ASV systems,
and Section~\ref{sec:con} concludes the paper.

\section{Trials and bias on trials}
\label{sec:bias}

In speaker verification, the performance of a system is evaluated by a set of trials.
Each trial is an individual test case, generally involving an enrollment prototype and a test speech.
With modern embedding-based systems (e.g., i-vector~\cite{dehak2011front} or x-vector~\cite{snyder2018xvector} systems),
both the enrollment prototype and the test speech are represented by embedding vectors,
and the verification task is formulated as testifying whether the two embedding vectors
belong to the same speaker or not.

Trials are used to measure performance of ASV models and systems, and the role is like a prober.
Obviously, if the prober is not well designed, the performance measurement cannot be reliable.
Unfortunately, the importance of trials have not widely recognized,
and the trivial \emph{cross-pairing} approach remains the most popular for trial design.
Specifically, suppose there are $N$ speakers in the test set, each having $K$ utterance,
{\bf full} cross-pairing selects any two utterances (or the corresponding embedding vectors) as a trial
and labels it as positive if the two utterances are from the same speaker or negative if they are from different speakers.
This leads to $NK(K-1)$ positive trials and $N(N-1)K^2$ negative trials. A variant of this full cross-pairing
is {\bf enrollment-fixed} cross-pairing, which fixes the enrollment speech, and selects only test speech by cross-pairing.
This leads to $NK$ positive trials and $N(N-1)K$ negative trials, where $K$ is the number of test utterances per speaker.

\begin{figure}[htb!]
\centering
\includegraphics[width=1.0\linewidth]{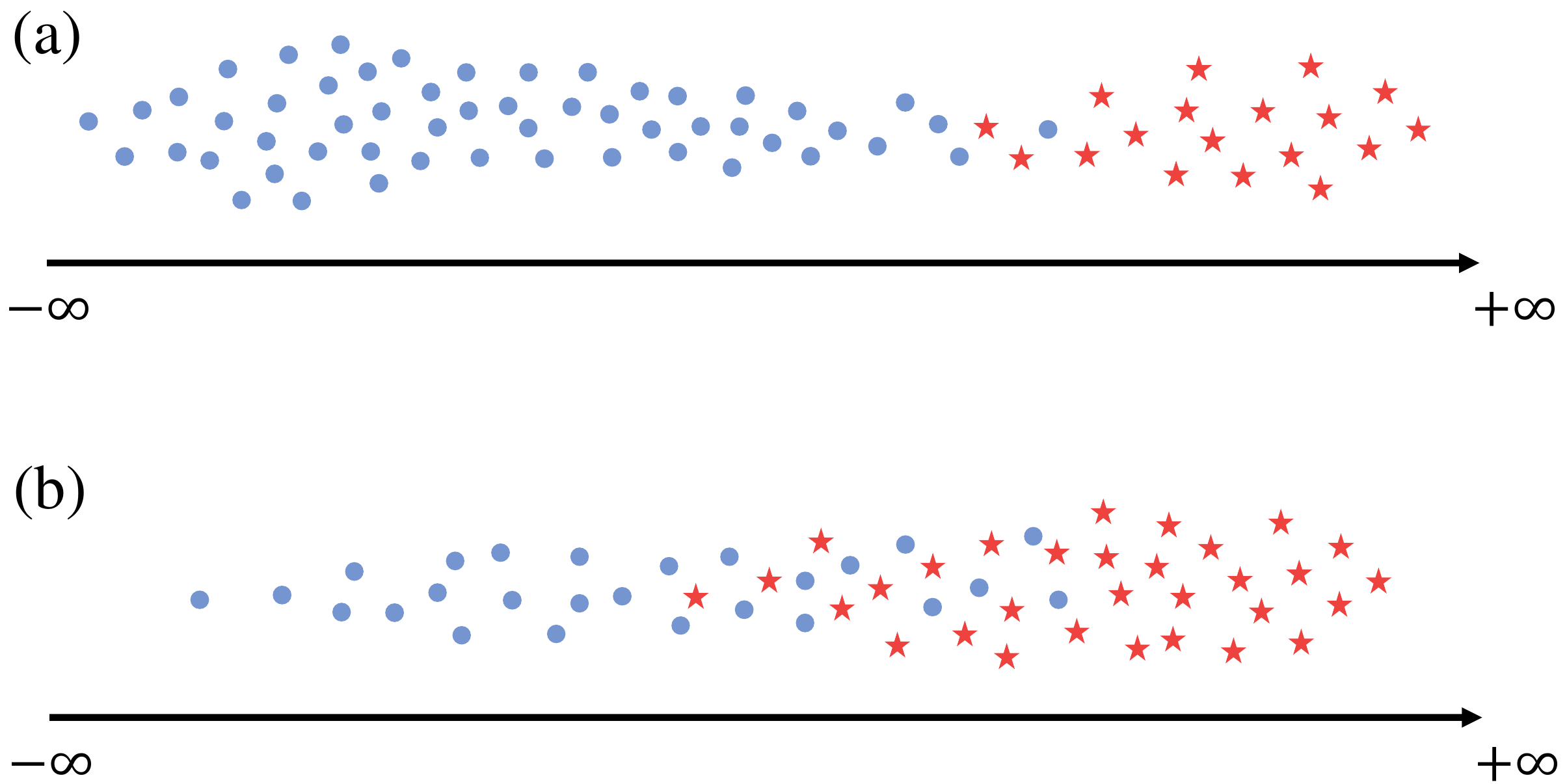}
\caption{Illustration of trial bias, where the axis represents the scores produced by an ASV system.
The top figure~(a) shows the scores of trials created by cross-pairing,
and the bottom figure~(b) shows the scores of trials encountered in a real application.
The red star represents positive trials and the blue circle represents negative trials.
Note that the class-conditional distributions of the scores are different in the two scenarios, which reflects the bias on trials.}
\label{fig:bias}
\end{figure}

A notable problem of the cross-pairing approach is that it produces a large proportion of \emph{easy trials}:
trials that are easy to make decisions. This is particularly the case for negative trials.
Unfortunately, the trials encountered in real-life applications might be quite different from
those created by the cross-pairing, and usually more challenging.
For instance, real situations tend to involve users with similar accent and utterances recorded in similar acoustic environment,
making the negative trials more challenging.
In addition, the number of positive trials could be much larger than negative trials, so the probability of meeting easy trials is low.
Finally, if a user finds a failure, he/she tends to try it over and over, leading to more hard trials.
All the above issues suggest that there is a \emph{trial bias} between the situations of benchmark test and real-life deployment.
We argue that trial bias is an important source for the benchmark-deployment gap
mentioned in the previous section. Fig.~\ref{fig:bias} illustrates the phenomenon of trial bias.

It should be highlighted that from the perspective of performance evaluation, the bias is only related to
the score distributions of the two trial classes (positive and negative).
This means (1) only the \emph{scores} matter, not any other properties of the test speech, such as acoustic environment,
speaker traits, etc. (2) only the \emph{distributions} matter, not any other quantities such as the number of speakers
and the utterances per speaker.
These two points are important and will be used to define the concept of trial config and C-P map.

\section{Trial config and C-P map}
\label{sec:map}

\subsection{Concept of C-P map}
\label{sec:cpmap}

To gain a deeper understanding for the bias on trials, we introduce the concept of \emph{trial config}.
Given a set of enrollment/test utterances,
a trial config is defined as \emph{a subset of trials selected to test against the target system}.
The full cross-pairing is the largest trial config and involves all the possible trials;
the enrollment-fixed cross-pairing is another trial config where utterances are separated into an enrollment set and a test set,
and only cross-set pairing is permitted.
Clearly, for a particular ASV system, performance with different trial configs are different,
reflecting performance of the target system under different deployment situations.
By collecting all the trial configs and the corresponding performance, we can evaluate the target system in a more thorough way.
This idea can be implemented as a \emph{config-performance (C-P) map}. In this map,
the x-axis corresponds to subsets of positive trials and the y-axis corresponds to subsets of negative trials,
so each location $(x,y)$ on the map corresponds to a particular trial config. Let the color
at $(x,y)$ represents the performance measurement, we obtain a C-P map.

\begin{figure}[htb!]
\centering
\includegraphics[width=0.90\linewidth]{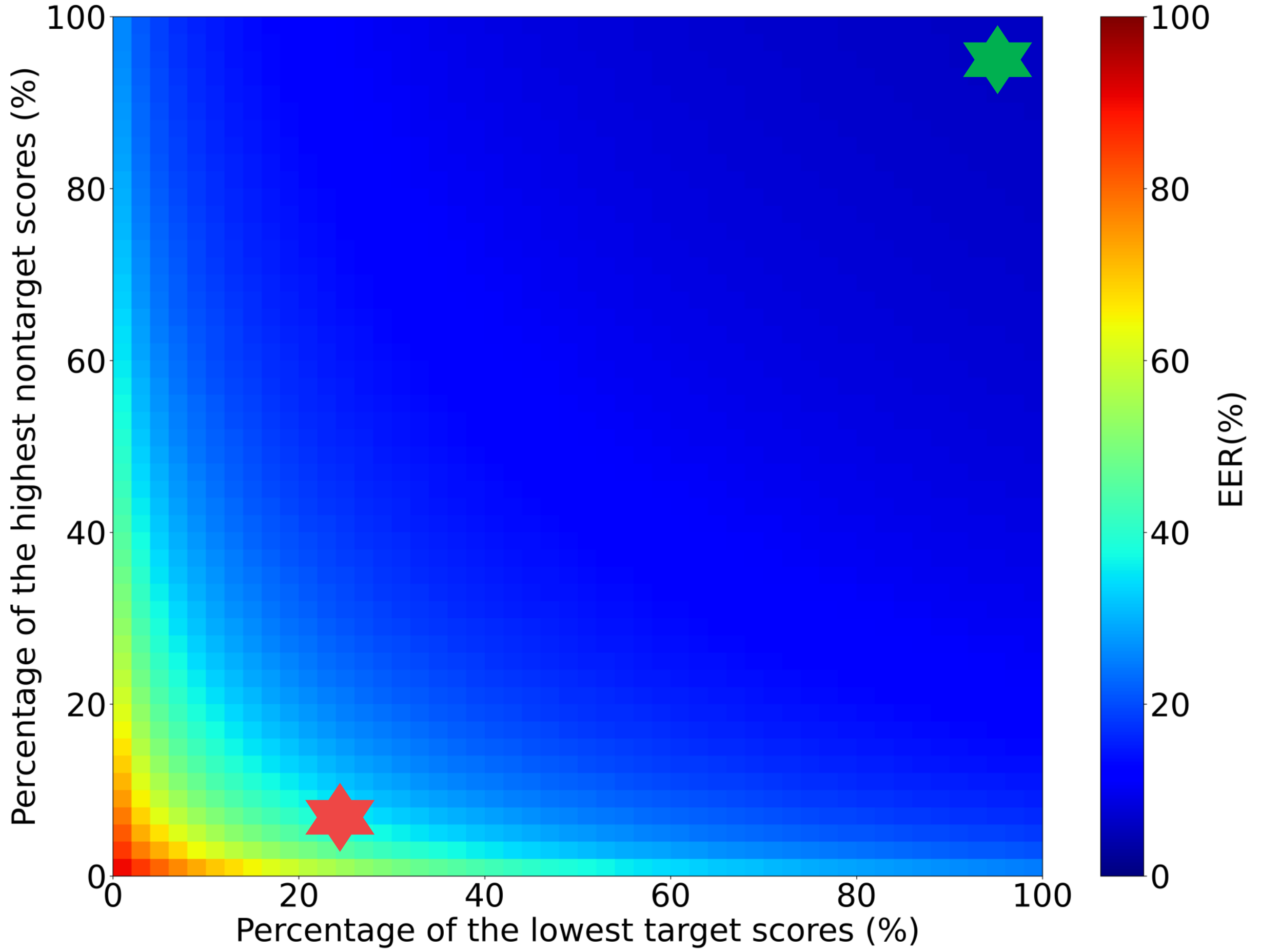}
\caption{Illustration of config-performance (C-P) map. The green star and the red start illustrates the
trial configs applied in benchmark test and real-life applications respectively.}
\label{fig:cp-map}
\end{figure}

A general C-P map is hard to read unless the neighboring trial configs form a spatial structure. A natural
structure can be obtained by sorting the trial configs according to their `hardness'. The hardness can be
defined by human, by agreement of several system, or by a system itself. As an example,
Fig.~\ref{fig:cp-map} shows a C-P map where the trial configs are sorted by the private hardness
of an i-vector system. Specifically, we sort the trials according to the scores assigned by an ASV system,
and then construct trial configs by selecting trials from the ordered list.
For the target trials sets (x-axis), we gradually select trials with higher scores from left to right,
and for the non-target trials sets (y-axis), we gradually select trials with lower scores from bottom to up.
The color in the map represents the EER values corresponding to each trial config.
This leads to an ordered C-P map, where the trial configs on the left and bottom are
harder than those on the right and up.

Although the hardness is defined privately by an ASV system, the C-P map still reveals lots of information.
For instance, the large proportion of the high-performance area (right and up) suggests that there are many easy trials in the
full cross-pairing trial config, and the low-performance area close to the origin point implies that the system performance is not perfect.
Moreover, the benchmark-deployment gap we mentioned in previous sections can be clearly explained with the C-P map.
First notice that the benchmark test and real-life deployment apply two different trial configs,
one is created with cross-pairing design and the other is determined by the application scenario.
These two configs are illustrated in Fig.~\ref{fig:cp-map} by the green star (right-up corner) and the
red star (on the bottom left) respectively. The position of the red star reflects the fact that
in real applications, negative trials could be more challenging,
and positive trials tend to be more than negative trials in number.
It is clear that the two trial configs lead to quite different EER results, which is precisely the benchmark-deployment gap.

If the order of the trial configs are fixed, the C-P map is more useful. For instance, it is possible to select
several ASV baseline systems to evaluate each trial, and use the averaged score to sort the trials and construct
the ordered trial configs. Using these ordered trial configs to draw C-P maps for different systems, we can get
detailed comparison among systems. This offers a powerful tool that can be used to confirm
effective techniques or identify spurious innovations.

\subsection{Validity of C-P map}

A critical question arises here is: are the performance measurements shown at different locations on the C-P map comparable, if they are
based on different sets of trials? If the answer is NO, then the C-P map is not very useful, as the patterns shown in the map will be meaningless.

\begin{figure}[htb!]
\centering
\includegraphics[width=\linewidth]{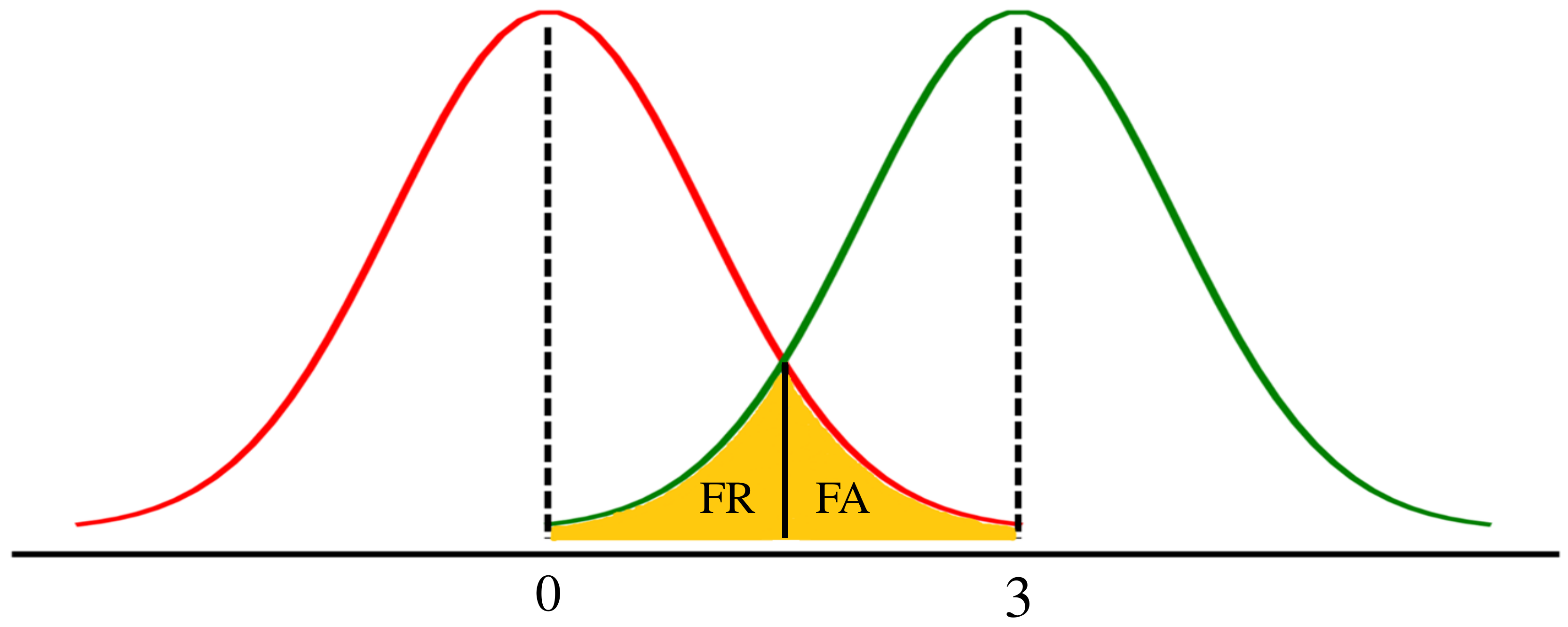}
\caption{Illustration of EER computation when the score distributions of the negative and positive trials are both Gaussian.
Suppose the variances of the two Gaussians are both 1.0, and the means are 0.0 and 3.0 respectively.
According to the symmetry, it is easy to know the decision threshold is 1.5, and looking the CDF table of a normal Gaussian shows that EER is 6.68\%.
}
\label{fig:gauss}
\end{figure}

To answer this question, firstly notice that the performance metrics we consider in this study, i.e, EER and minDCF,
are fully determined by the score distributions of the positive and negative trial classes.
Take EER as an example, if we use $p(c)$ and $q(c)$ to denote the score distributions of positive and negative trials respectively,
then it is easy to compute the EER threshold $\theta$ by solving the following equation:

\begin{equation}
\label{eq:eer}
\int_{-\infty}^\theta p(c) {\rm d}c = \int_{\theta}^{+\infty} q(c) {\rm d}c
\end{equation}

\noindent where the left quantity represents the false rejection (FR) rate and the right quantity represents the false alarm (FA) rate,
when the decision threshold is set to $\theta$. According to the monotonic property of cumulative distribution functions (CDFs),
the solution for $\theta$ is unique. Once $\theta$ is determined, the EER can computed as the left or right quantity of Eq.~(\ref{eq:eer}).
Fig.~\ref{fig:gauss} shows an example of the computation
when the score distributions of both the positive and negative classes are Gaussian.

This means that no matter which trials are used in the test set,
only if they can approximate the score distributions of the positive and negative classes,
the EER/minDCF results will approach to the true values.
For instance, if one increases the number of test speakers and the number of test utterances per speaker,
the number of negative trials will be increased much faster than the number of positive trials,
but the EER/minDCF will be the same, supposing that the new speakers and utterances are sampled from the same
underlying distribution as the original test set.

Since the evaluation metrics are determined by distributions of scores of trials rather than trials themselves,
it is obvious to conclude that the performance values in the C-P map are comparable, only if every trial config
involves sufficient trials. Note that the condition of the above statement means that on the C-P map,
the origin point and its vicinity are not well defined and should be interpreted with caution.

\section{Experiments}
\label{sec:exp}

\subsection{Data}
\label{sec:data}

VoxCeleb~\cite{nagrani2017voxceleb,chung2018voxceleb2} was used in our experiments.
It is a large-scale audio-visual speaker dataset collected by the University of Oxford, UK\footnote{http://www.robots.ox.ac.uk/$\sim$vgg/data/voxceleb/}.
Specifically, the development set of \emph{VoxCeleb2} was used to train the i-vector and x-vector systems,
which contains 5,994 speakers in total and entirely disjoints from the VoxCeleb1 and SITW datasets.
Trials of the cleaned \emph{VoxCeleb1-O} was used for tuning and the cleaned \emph{VoxCeleb1-E} was used for evaluation.

\begin{table*}[htb!]
\centering
\caption{EER(\%) and minDCF results with the modern ASV systems on VoxCeleb1 evaluation trials.}
\vspace{1mm}
\label{tab:base}
\begin{tabular}{cllcccc}
\toprule
\multirow{2}{*}{System} & \multirow{2}{*}{Front-End}   &  \multirow{2}{*}{Back-End}  & \multicolumn{2}{c}{VoxCeleb1-O}  & \multicolumn{2}{c}{VoxCeleb1-E}  \\
                     &                                 &                             & EER(\%)    & minDCF              & EER(\%)    & minDCF~~        \\
\cmidrule(r){1-7}
1                    &   GMM i-vector                  &   PLDA                      & 5.819      & 0.5189              & 5.872      & 0.5038          \\
\cmidrule(r){1-7}
2                    &   TDNN + TSP + Softmax          &   PLDA                      & 4.558      & 0.4882              & 4.290      & 0.4343          \\
\cmidrule(r){1-7}
3                    &   TDNN + TSP + AM-Softmax       &   Cosine                    & 3.430      & 0.3370              & 3.389      & 0.3619          \\
\cmidrule(r){1-7}
4                    &   ResNet34 + TSP + AM-Softmax   &   Cosine                    & 1.633      & 0.1770              & 1.688      & 0.1900          \\
\cmidrule(r){1-7}
5                    &   ResNet34 + TSP + AAM-Softmax  &   Cosine                    & 1.803      & 0.1961              & 1.747      & 0.1946          \\
\cmidrule(r){1-7}
6                    &   ResNet34 + ASP + AM-Softmax   &   Cosine                    & 1.521      & 0.1642              & 1.504      & 0.1669          \\
\bottomrule
\end{tabular}
\end{table*}

\subsection{Basic systems}
\label{sec:base}

In our experiment, we firstly follow the VoxCeleb recipe of the Kaldi toolkit~\cite{povey2011kaldi}
to build our i-vector~\cite{dehak2011front} and x-vector~\cite{snyder2018xvector} baselines\footnote{https://github.com/kaldi-asr/kaldi/tree/master/egs/voxceleb/}.
These basic recipes may not achieve the best performance on a particular dataset,
but have been demonstrated to be highly competitive and generalizable by many researchers with their own data and model settings.
Moreover, using these recipes allow others to reproduce our results easily. No data augmentation is used.

\begin{itemize}

\item i-vector: The i-vector model was built following the Kaldi VoxCeleb/v1 recipe.
The acoustic features comprise 23-dimensional MFCCs plus the log energy, augmented by the first- and second-order derivatives,
resulting in a 72-dimensional feature vector.
Moreover, cepstral mean normalization (CMN) is employed to normalize the channel effect,
and an energy-based voice active detection (VAD) is used to remove silence segments.
The UBM consists of 2,048 Gaussian components, and the dimensionality of the i-vector space is 400.
For the back-end model, LDA is firstly used to reduce dimensionality to 200, and then PLDA~\cite{Ioffe06} is employed to score the trials.

\item x-vector: The x-vector model was created following the Kaldi VoxCeleb/v2 recipe.
The acoustic features are 30-dimensional MFCCs.
The DNN architecture adopts TDNN topology which involves 5 time-delay (TD) layers to learn frame-level deep speaker features,
and a statistic pooling (TSP) layer is used to accumulate the frame-level features to utterance-level statistics,
including the mean and standard deviation.
After the pooling layer, 2 fully-connection (FC) layers are used as the classifier, for which the outputs correspond to
the number of speakers in the training set.
Once trained, the 512-dimensional activations of the penultimate layer are read out as an x-vector.
The back-end model is the same as in the i-vector system.

\end{itemize}

\subsection{More powerful systems}
\label{sec:sota}

We also constructed more powerful x-vector systems to reproduce the performance of the SOTA speaker recognition techniques.
A bunch of state-of-the-art architectures/techniques, including ResNet34 topology~\cite{he2016deep},
attentive statistics pooling (ASP) strategy~\cite{okabe2018attentive}
and angular margin-based training objectives (AM-Softmax~\cite{wang2018additive} and AAM-Softmax~\cite{deng2019arcface})
are employed in our experiments.
Specifically, the acoustic features are 80-dimensional Fbanks.
The DNN architecture adopts the ResNet34 topology for frame-level feature extraction.
The TSP and ASP strategies are used to construct utterance-level representations.
These representations are then transformed by a
fully-connected layer to generate logits and are fed to a softmax layer to generate posterior probabilities over speakers.
The training objective employs AM-Softmax and AAM-Softmax, and the margin factor is set to 0.2 and the scale factor is set to 30.
Once trained, the 256-dimensional activations of the last fully-connection layer are read out as an x-vector.
The simple cosine distance is used to score the trials.
No data augmentation is used.
The source code has been published online to help readers reproduce our systems\footnote{https://gitlab.com/csltstu/sunine}.

~\\
The EER and minDCF (p-target is 0.01) results of these ASV systems are presented in Table~\ref{tab:base}.
It can be observed that modern ASV systems can obtain reasonable performance, even with the basic settings.

\subsection{C-P map}
\label{sec:cp}

In this section, we use the C-P map as a tool to analyze performance of the modern ASV systems on the cleaned \emph{VoxCeleb1-E} dataset.
In order to define a unified arrangement for the trial configs in the C-P map,
the scores from the two basic systems (i-vector and x-vector) are averaged and used to sort the positive and negative trials respectively.
The C-P maps are then constructed based on the sorted trials, following the process presented in Section~\ref{sec:cpmap}.

\begin{figure*}[htb!]
\centering
\includegraphics[width=0.96\linewidth]{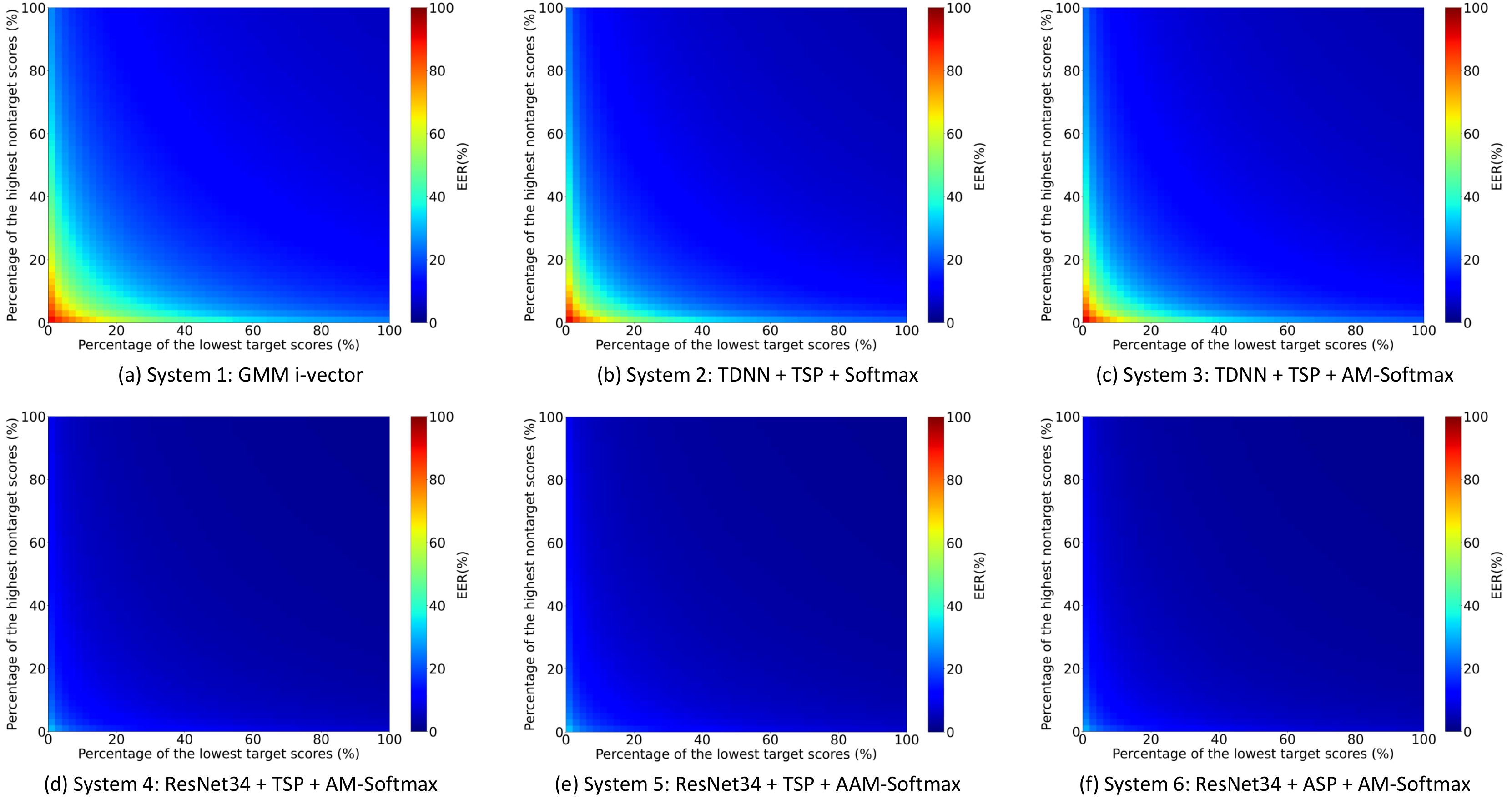}
\caption{The C-P maps of 6 systems tested on VoxCeleb-E trials with \emph{EER} metric.}
\label{fig:cpmap-eer}
\end{figure*}

\begin{figure*}[htb!]
\centering
\includegraphics[width=0.96\linewidth]{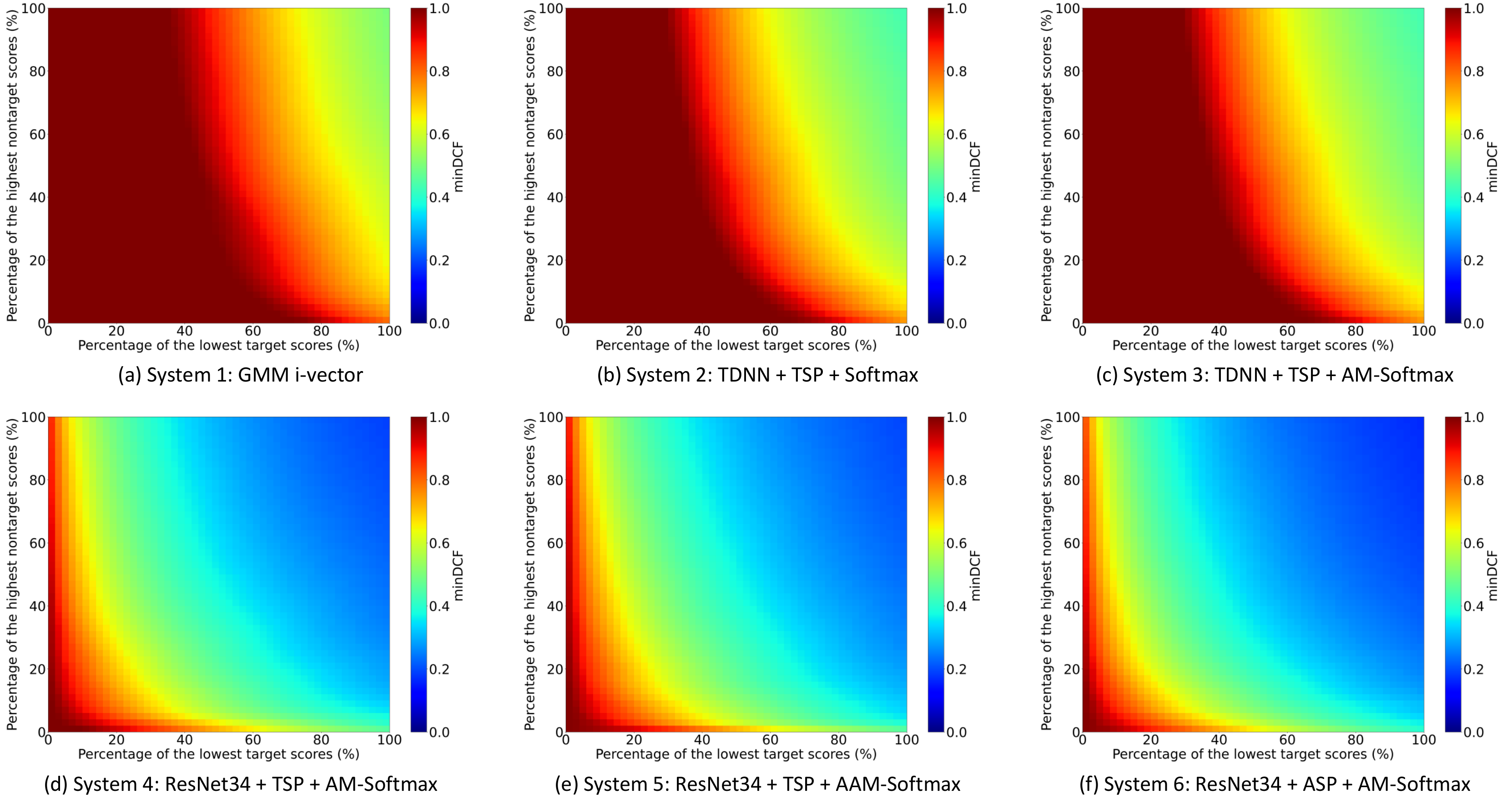}
\caption{The C-P maps of 6 systems tested on VoxCeleb-E trials with \emph{minDCF} metric.}
\label{fig:cpmap-dcf}
\end{figure*}

Fig.~\ref{fig:cpmap-eer} and Fig.~\ref{fig:cpmap-dcf} show the C-P maps of 6 systems corresponding to Table~\ref{tab:base} with EER and minDCF metrics respectively.
Firstly, it can be seen that all these systems can obtain satisfactory performance on the trial configs in the right-up corner.
However, as the trial configs approach to the origin point in the left-bottom corner, the performance gap of different systems becomes significant.
This indicates that these systems are good at recognizing easy trials but are different on tackling hard trials.
Secondly, comparing these C-P maps, we observe that with the development of speaker recognition techniques (from System 1 to System 6), the performance improvement is steadily obtained.
Among all these technique advances, neural topology change from TDNN to ResNet34 seems the most significant.

\subsection{Delta C-P map}
\label{sec:delta-cp}

This section uses C-P map to compare two systems on various trial configs.
To make the comparison more clear, the relative change ratio (RCR) at each location $(x,y)$ on two C-P maps is computed as follows:

\begin{equation}
\label{eq:rcr}
  \text{RCR}(x,y) = \frac{\text{CP}_{ref}(x, y) - \text{CP}_{test}(x, y)}{\text{CP}_{ref}(x, y)},
\end{equation}
\noindent where $\text{CP}_i(x,y)$ is the metric value (e.g., EER, minDCF) at the location $(x,y)$ on the C-P map of system $i$.
$\text{CP}_{ref}$ represents the C-P map of the reference system and $\text{CP}_{test}$ represents the C-P map of the test system.

Once $\text{RCR}(x,y)$ is computed at all the locations, we can construct a \emph{delta C-P map}, to compare
two systems in an all-round way.
Obviously, for the trial config at location $(x,y)$ in the delta C-P map, if its $\text{RCR}$ is positive, it means the test system 
beats the reference system. In contrast, if $\text{RCR}(x,y)$ is negative, the test system loses.
To avoid numerical uncertainty, we define a small tolerance value $\epsilon$ ($10^{-5}$ in this study), 
and regard the comparison is tied if $|\text{RCR}(x,y)| < \epsilon$.
We can therefore compute the proportion of the trial configs on the delta C-P map with three situations (\emph{win: tie: lose}),
which provides a thorough comparison between two systems.
Experiments are conducted to compare some representative speaker recognition techniques by delta C-P map,
from the perspectives of model, topology, training objective and pooling strategy.
All these results demonstrate that delta C-P map is a powerful tool to compare performance of systems in a thorough way.

\begin{figure*}[htb]
\centering
\includegraphics[width=0.79\linewidth]{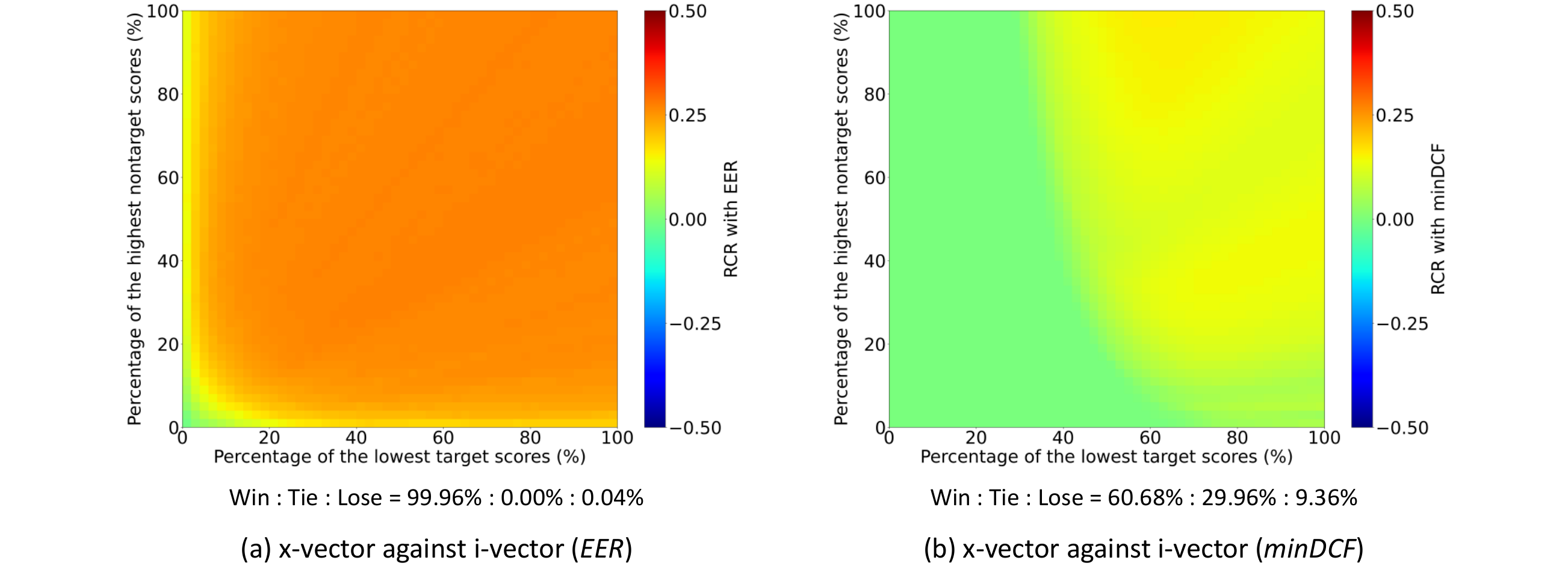}
\vspace{-1.5mm}
\caption{The delta C-P maps of x-vector against i-vector with \emph{EER/minDCF} metrics.}
\label{fig:delta-cp-vec}
\end{figure*}

\begin{figure*}[htb]
\centering
\includegraphics[width=0.79\linewidth]{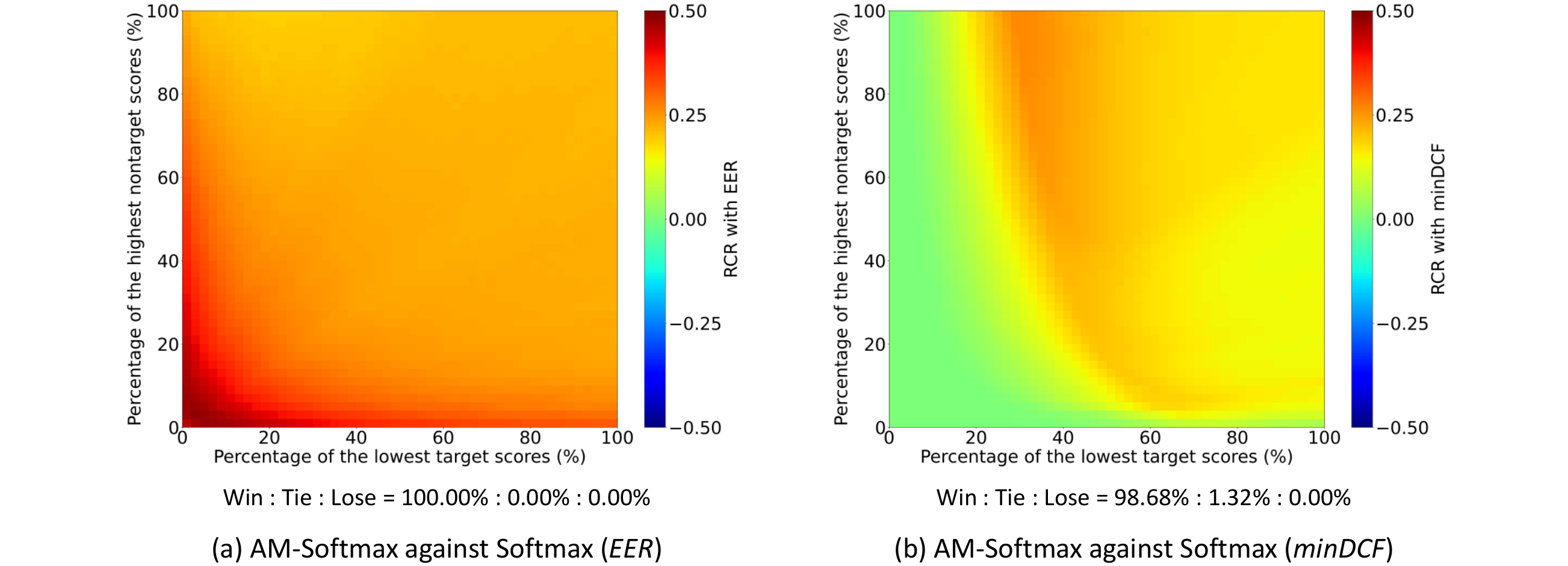}
\vspace{-1.5mm}
\caption{The delta C-P maps of AM-Softmax against Softmax with \emph{EER/minDCF} metrics.}
\label{fig:delta-cp-soft}
\end{figure*}

\begin{figure*}[htb!]
\centering
\includegraphics[width=0.79\linewidth]{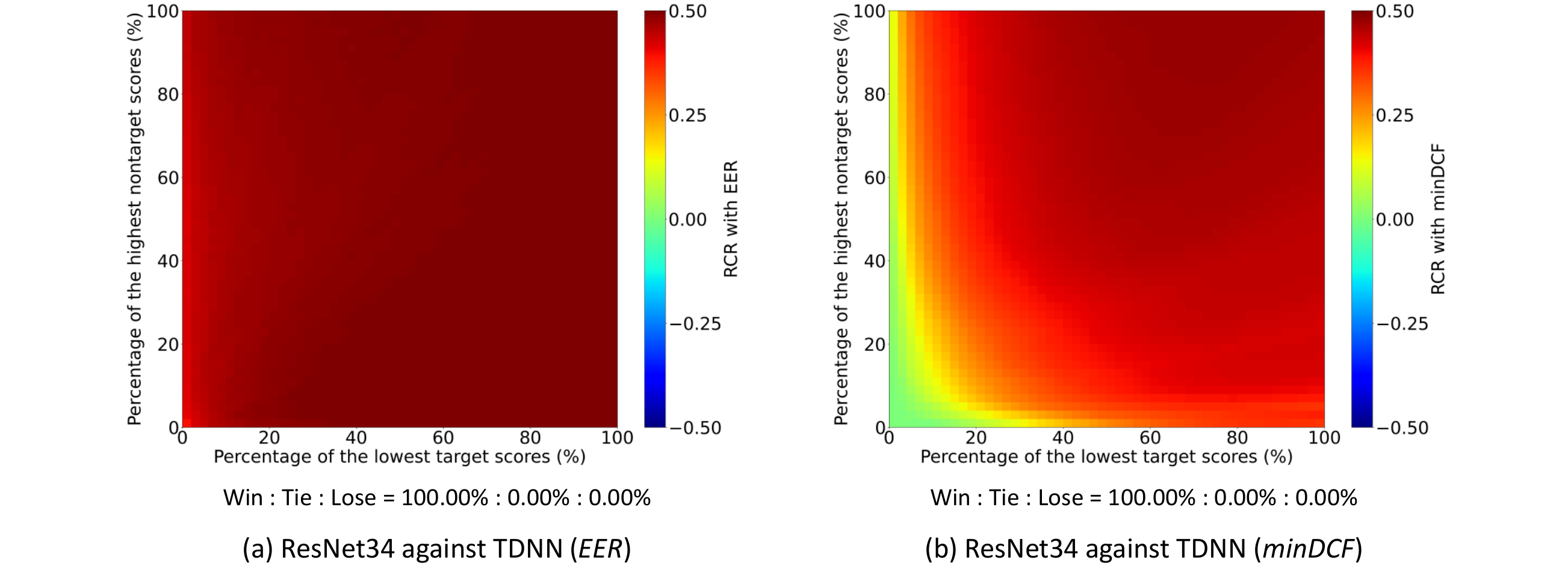}
\vspace{-1.5mm}
\caption{The delta C-P maps of ResNet34 against TDNN with \emph{EER/minDCF} metrics.}
\label{fig:delta-cp-net}
\end{figure*}

\begin{figure*}[htb!]
\centering
\includegraphics[width=0.79\linewidth]{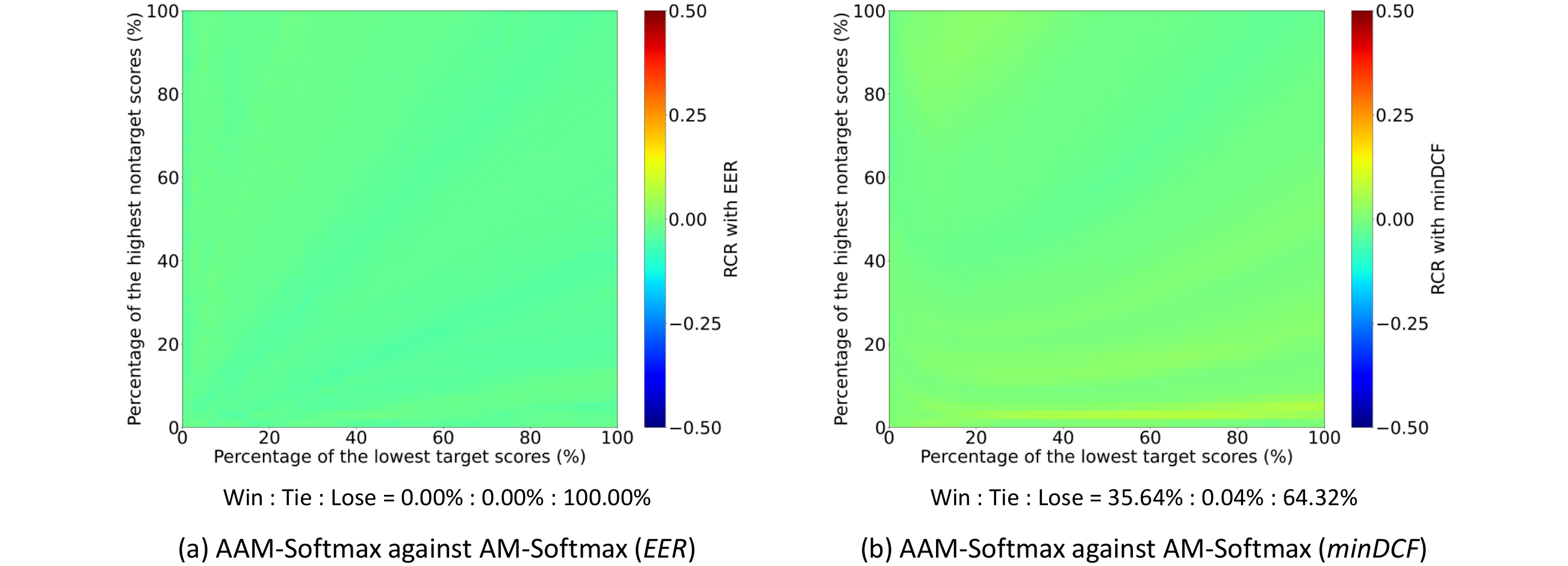}
\vspace{-1.5mm}
\caption{The delta C-P maps of AAM-Softmax against AM-Softmax with \emph{EER/minDCF} metrics.}
\label{fig:delta-cp-obj}
\end{figure*}

\begin{figure*}[htb!]
\centering
\includegraphics[width=0.79\linewidth]{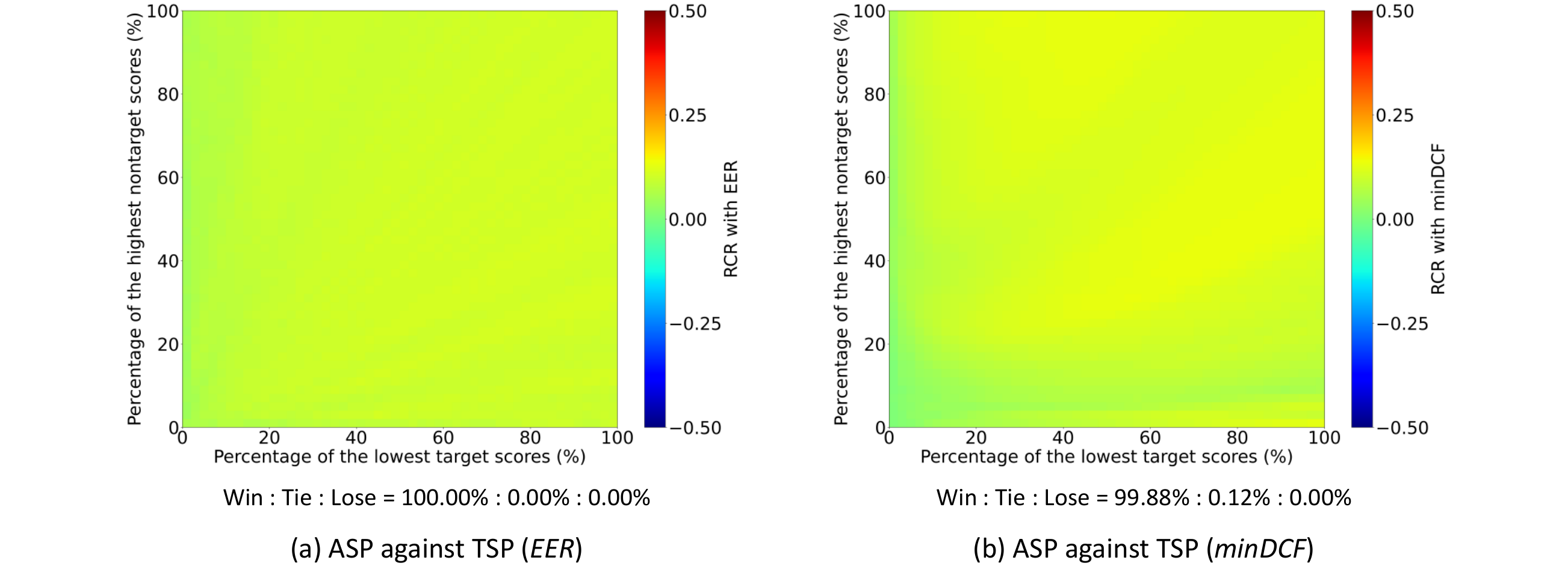}
\vspace{-1.5mm}
\caption{The delta C-P maps of ASP against TSP with \emph{EER/minDCF} metrics.}
\label{fig:delta-cp-pool}
\end{figure*}

\begin{figure*}[htb!]
\centering
\includegraphics[width=1.0\linewidth]{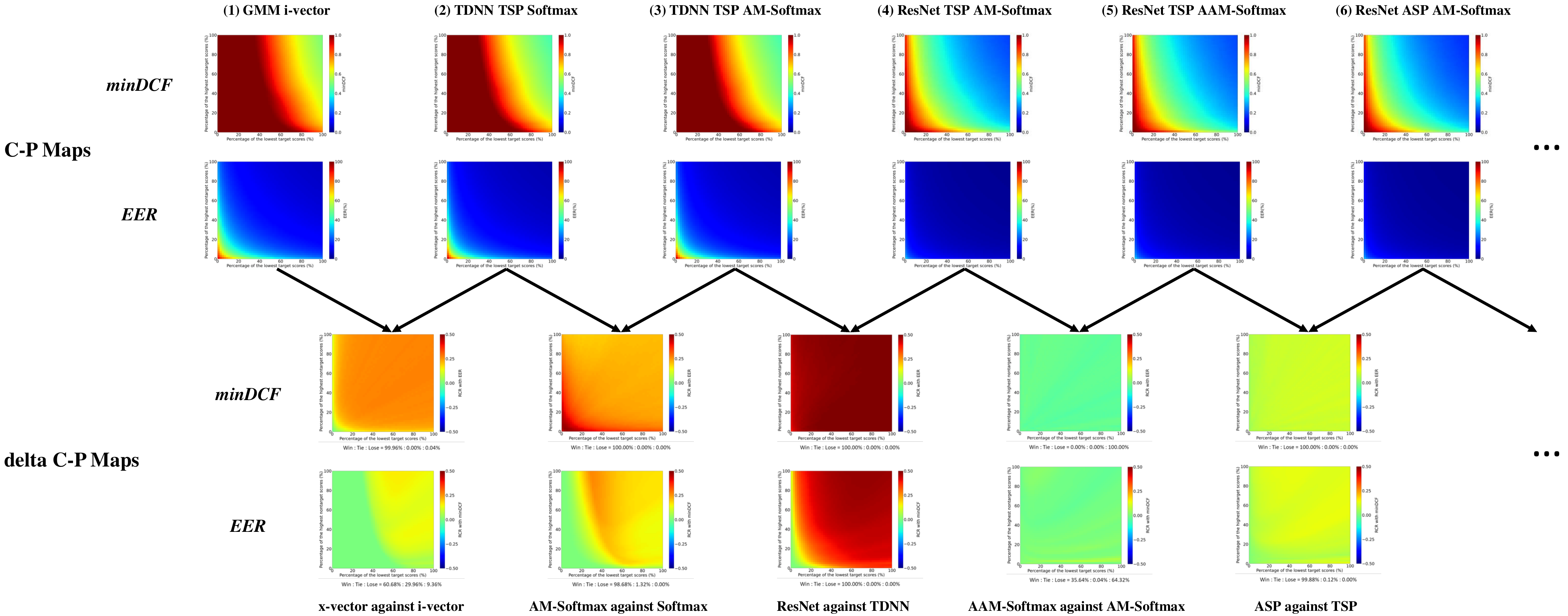}
\caption{The roadmap of speaker recognition techniques measured by C-P map and delta C-P map.}
\vspace{-1.5mm}
\label{fig:roadmap}
\end{figure*}

\subsubsection{i-vector vs. x-vector}

Firstly, the delta C-P map between the basic i-vector system  and x-vector system is presented, as shown in Fig.~\ref{fig:delta-cp-vec}.
The C-P map of the i-vector system (System 1) corresponds to $\text{CP}_{ref}(\cdot)$ and the C-P map of the x-vector system (System 2) corresponds to $\text{CP}_{test}(\cdot)$.
It can be observed that x-vector wins the game on almost all the trial configs.
This indicates that the discriminative model is superior to the probabilistic model at least under our experimental condition.

\subsubsection{Softmax vs. AM-Softmax}

Secondly, we compared two training objectives Softmax and AM-Softmax by delta C-P map, as shown in Fig.~\ref{fig:delta-cp-soft}.
The C-P map of the Softmax system (System 2) corresponds to $\text{CP}_{ref}(\cdot)$ and the C-P map of the 
AM-Softmax system (System 3) corresponds to $\text{CP}_{test}(\cdot)$.
It can be observed that the margin-based AM-Softmax overwhelmingly outperforms the standard Softmax, even though with a simpler back-end model.

\subsubsection{TDNN vs. ResNet34}

Thirdly, we compare two popular neural topologies, TDNN  and ResNet34 by delta C-P map, as shown in Fig.~\ref{fig:delta-cp-net}.
The C-P map of the TDNN system (System 3) corresponds to $\text{CP}_{ref}(\cdot)$ and the C-P map of the ResNet34 system (System 4) corresponds to $\text{CP}_{test}(\cdot)$.
It is clear to see that the ResNet34 topology is superior to TDNN by a large margin, demonstrating the success of ResNet34 in speaker recognition.

\subsubsection{AM-Softmax vs. AAM-Softmax}

Fourthly, two margin-based training objectives AM-Softmax and AAM-Softmax are compared by delta C-P map, as shown in Fig.~\ref{fig:delta-cp-obj}.
The C-P map of the AM-Softmax (System 4) corresponds to $\text{CP}_{ref}(\cdot)$ and the C-P map of AAM-Softmax (System 5) corresponds to $\text{CP}_{test}(\cdot)$.
It can be observed that the performance gap between the two objectives are quite marginal, even though AM-Softmax shows a bit advantage than AAM-Softmax on most trial configs.

\subsubsection{TSP vs. ASP}

Finally, two pooling strategies TSP and ASP are compared by delta C-P map, as shown in Fig.~\ref{fig:delta-cp-pool}.
The C-P map of the TSP system (System 4) corresponds to $\text{CP}_{ref}(\cdot)$ and the C-P map of the ASP system (System 6) corresponds to $\text{CP}_{test}(\cdot)$.
It can be found that ASP outperforms TSP on the whole, demonstrating its advantage.

\subsection{Roadmap}

This section combines all these C-P maps and delta C-P maps in Section~\ref{sec:cp} and Section~\ref{sec:delta-cp},
and constructs a roadmap that illustrates the development of speaker recognition techniques in recent years.
By this roadmap, it is clear to see which technique is effective and which innovation is revolutionary.
This further demonstrates that the proposed C-P map is a very valuable tool for technique analysis and system comparison.

\section{Conclusions}
\label{sec:con}

This paper is inspired by the ubiquitous benchmark-deployment discrepancy.
Our  hypothesis is that this problem is largely attributed to the different trials encountered in different situations.
To formulate the significant impact of trials on performance measurement, we propose the concept of C-P map which represents performance of a speaker verification system on
various trials configs in a 2-dimensional map.
We empirically show that C-P map is a novel evaluation toolkit for ASV system analysis and comparison, and can be used to confirm really effective techniques.
As for the future work, more comprehensive analysis will be conducted to understand the behavior of different models at different locations on the C-P map. 
Besides, other recently proposed techniques will be analyzed by C-P map to verify their effectiveness.

\newpage

\bibliographystyle{IEEEbib}
\bibliography{refs}

\end{document}